\documentclass[11pt]{article}
\usepackage[margin=1.4in]{geometry}
\usepackage{amsmath, amssymb, amsthm}
\usepackage{graphicx}
\usepackage{subfigure}
\usepackage{multirow}
\usepackage{booktabs}
\usepackage{algorithm}
\usepackage{algpseudocode}
\usepackage{siunitx}
\usepackage{url}
\usepackage[colorlinks=true,allcolors=black]{hyperref}
\usepackage[utf8]{inputenc}
\usepackage[T1]{fontenc}
\usepackage{lmodern}
\usepackage[english]{babel}
\usepackage{microtype}
\begin{document}

\newgeometry{left=1.5in, right=1.5in, top=2in, bottom=2in}

\begin{center}

{\LARGE\bfseries Status report towards implementation of a Compton polarimeter at BEPC\uppercase\expandafter{\romannumeral 2}}\\[1.8em]

\large
Mengyu Su$^{1,2}$,
Zhe Duan$^{1,3,\ast}$,
Jie Gao$^{1,2,3}$,
Qingfu Han$^{1}$,\\
Daheng Ji$^{1}$,
Dazhang Li$^{1}$,
Qi Li$^{1}$,
Xiaodong Li$^{1}$,
Yanchun Li$^{1}$,\\
Zhijun Liang$^{1,\ast}$,
Ge Lei$^{1}$,
Zhonghui Ma$^{1}$,
Aur\'{e}lien Martens$^{4,\ast}$,\\
Lingling Men$^{1}$,
Xianjing Sun$^{1}$,
Guangyi Tang$^{1}$,
Huan Wang$^{1}$,\\
Jianli Wang$^{1}$,
Lin Wang$^{1}$,
Qi Yang$^{1}$,
Lingda Yu$^{1}$,
Chenghui Yu$^{1,2,\ast}$ \\
Jianyong Zhang$^{1}$,
Wan Zhang$^{1}$,
Yandong Zhang$^{1}$,
Yuelei Zhang$^{1}$,\\
Ningchuang Zhou$^{1}$,
Dechong Zhu$^{1}$,
Fabian Zomer$^{4}$ \\[0.8em]

\large
$^1$ Institute of High Energy Physics, Chinese Academy of Sciences, 19B Yuquan Road, Shijingshan District, Beijing, 100049, China. \\
$^2$ University of Chinese Academy of Sciences, 19A Yuquan Road, Shijingshan District, Beijing, 100049, China. \\
$^3$ Center for High Energy Physics, Henan Academy of Sciences, 228 Chongshili, Zhengdong New District, Zhengzhou, 450046, China. \\
$^4$ Universit\'{e} Paris-Saclay, CNRS/IN2P3, IJCLab, Orsay, 91405, France. \\[1.8em]

$^\ast$ Corresponding author(s). E-mail(s): duanz@ihep.ac.cn; liangzj@ihep.ac.cn; aurelien.martens@ijclab.in2p3.fr; yuch@ihep.ac.cn; 

\vspace{2.5em}
\end{center}

\begin{abstract}
Precision beam polarization measurements based on Compton polarimeters are essential for the physics program of future high-energy colliders. In order to prepare for these and to extend the scope of physics measurements of the BES\uppercase\expandafter{\romannumeral 3} experiment at the BEPC\uppercase\expandafter{\romannumeral 2}, a diagnostic of electron beam transverse polarization at BEPC\uppercase\expandafter{\romannumeral 2} is of interest. The design and status report of the commissioning, until July 2025, of this device is reported in this paper. We report unambiguous observation of Compton interaction, discuss current limitations of the experimental setup and draw prospects for improvements and actual measurement of electron beam polarization in the near future.
\end{abstract}


\clearpage  
\restoregeometry

\section{Introduction}

Polarized lepton beams are essential for the physics program of future colliders~\cite{Brunner:CLIC_2022, Behnke:ILC_TDR_2013, Willeke:eiccdr_2021,FCCee_Feasibility_Study}, 
such as the Circular Electron Positron Collider (CEPC)~\cite{cepc_TDR_2024}. On one hand, the resonant depolarization~\cite{Bukin:1975db,
RD-Derbenev:1980} of transversely polarized
beam is essential for precision beam energy calibration at the Z-pole and the WW threshold, 
and, in particular, necessary for precision measurements of masses of fundamental particles~\cite{LEPCalib,Blondel:FCC_polairzation_and_energy_2019,xia_evaluation_2023}. 
This technology provides energy calibration with a relative accuracy of $10^{-6}$ as demonstrated at VEPP-4M~\cite{blinovResonantDepolarizationTechnique2022}.
On the other hand, the collision of longitudinally polarized beams enhances the physics reach of 
experiments for a given integrated luminosity~\cite{CLENDENIN19941,Moortgat-Pick:2008,USBelleIIGroup:SKEKB2022qro}. 
For CEPC,
a scheme of generating polarized beams from the source and injecting
them into the collider rings has been studied~\cite{Duan:2023cgu,chen_booster_2023}, implementation of spin rotators in the collider ring
facilitates longitudinal 
polarization of electron beams at Z-pole~\cite{Xia:spinrotator}. 

Polarization measurement is the basis of these polarized beam studies. 
Two techniques are commonly employed to measure polarization of high energy beams. On the one hand, Møller scattering can be exploited either by implementing a dedicated polarimeter, destructive for the beam~\cite{King:2023zbf-JLab} or non-destructively by disentangling and monitoring the Touschek contribution to the beam lifetime~\cite{blinovReviewBeamEnergy2009a, PhysRevAccelBeams.22.122801, SUN2010339, Fu:2024mhq-Touscheklifetime}. However, the former cannot be employed in a ring collider for real time polarization measurements, the latter essentially provides information on the magnitude of the polarization vector without the orientation.
On the other hand, Compton polarimetry~\cite{baierkhoze} has been widely used
\cite{PhysRevLett.66.1697,Gustavson1979, FRANKLIN200061, BARBER199379,sobloher2012polarizationherareanalysis,CBartels_2012,IMAI2005332, BECKMANN2002334,JLabcompton_2005, Baudrand_2010, hillert_compton_2009, placidiPolarimetryLEP1989a,Blinov_2017,Kaminskiy:VEPP2024,Woods:1996nz} to provide non-destructive polarization measurements, possibly in real-time~\cite{Charlet_2023} and potentially sensitive for detecting all polarization components~\cite{Muchnoi_2022}. 
Recently a relative accuracy of $dP/P = 0.36\%$ has been obtained~\cite{PhysRevC.109.024323} for longitudinal polarization measurements, and about $1.9\%$ for transverse polarization measurements~\cite{sobloher2012polarizationherareanalysis}. 

The differential cross section for Compton scattering of a photon of energy $\hbar\omega_0$ off an electron of energy $E$ reads~\cite{BARBER199379,Berestetskii:1982qgu,ginzburg_colliding_1984,Charlet_2023}
\begin{equation}
\frac{d^2\sigma}{du d\phi}=\frac{d\sigma_0}{du}\left[1+f_1(\kappa,u,S_1,S_2,\phi)+S_3\left(P_za_z(\kappa,u)+P_ta_t(\kappa,u)\cos\phi\right)\right],
\label{equ:dif_cross_section}
\end{equation}
where $\kappa=\frac{2E\hbar\omega_0(1+\beta\cos\theta)}{m_e^2c^4}$, $u=\hbar\omega/E$,
$\hbar\omega$ is the energy of a scattered photon, $m_e$ is the rest mass of an electron, $\phi$ is the azimuthal angle of the direction of the outgoing photon with respect to the electron transverse polarization, and 
$\frac{d\sigma_0}{du}$ is the unpolarized differential cross-section. $S1, S2, S3$ are Stokes parameters normalized to the total laser intensity, 
corresponding to linear horizontal/vertical, linear ${\pm}45 ^{\circ}$ orientation and circular polarization, respectively. 
$P_z$ and $P_t$ are the longitudinal and transverse components of electron polarization, while $a_z$ and $a_t$ are the 
corresponding coefficients of these different components.
It is clear that the detection of electron beam polarization
relies on extracting the term proportional to the degree of circular
polarization of the laser beam $S_3$. Thus, in Compton polarimetry experiments, the laser beam is typically circularly polarized. The determination of a related systematic uncertainty is subject to dedicated studies when very high precision is sought for~\cite{Brisson_2010, PhysRevC.109.024323}. The function $f_1$ represents the dependence of the cross-section on the linear polarization of the laser, which is thus expected to be small 
due to the dominant circular polarization.
Expressions for the functions $\frac{d\sigma_0}{du}$, $f_1$, $a_{z,t}$ can be obtained from the literature~\cite{ginzburg_colliding_1984,Charlet_2023}, and are not reproduced here for brevity.

Furthermore, the measurement of the energy distribution of scattered photons (and conversely electrons) provides a measurement of the 
electron\slash positron
beam longitudinal component of the polarization. 
Sensitivity to the transverse component, of interest for resonant depolarization experiments, is obtained only when the azimuthal distribution of the scattered particles is probed. As a consequence, transverse polarization is notoriously more difficult to measure since the direction of photons must be disentangled within the $1/\gamma$ opening cone, where $\gamma=E/m_ec^2$ is the Lorentz boost of initial electrons. Except for the recent VEPP-4M implementation~\cite{Blinov_2017} and recent result at ELSA~\cite{switka:ibic2025-wepmo32}, transverse electron beam polarimetry experiments relied mainly on scarce vertical sampling of detectors~\cite{hillert_compton_2009, Assmann-LEPpolarimeter, BARBER199379,BECKMANN2002334}.
In order to reach the needed precision for future colliders, silicon pixel detectors will likely be implemented and operated for the measurement of high energy electrons/positrons and photons~\cite{Mordechai:2013zwm,Muchnoi_2022,CEPC_Polarimeter,switka:ibic2025-wepmo32}.
This technique is very promising, as it might allow for a precision measurement of all electron beam polarization parameters as well as providing information on the nuisance parameters related to not perfectly controlled residual ellipticity in the laser beam polarization~\cite{Muchnoi_2022}. It also requires the implementation of a simultaneous measurement of the scattered photons and electrons azimuthal distribution. As a consequence, in view of future operation of Z, W and H-factories, the implementation of a pixel-based Compton polarimeter at a lower energy but existing collider is of interest. In order to progress towards proof of concept, we aim at implementing a pixel sensor for the detection of photons, whose working principles remain close to the most recent VEPP-4M implementation~\cite{Blinov_2017}. This proof of concept has been designed~\cite{su_preliminary_2024} and implemented in the electron storage ring of BEPC\uppercase\expandafter{\romannumeral 2}~\cite{Yu:IPAC2016cof}.

Towards reaching this goal, this paper reports on the design, implementation and first beam commissioning of this Compton polarimeter at BEPCII, but with a simplified detector to unambiguously assess the detection of Compton scattering. The implementation of the pixel detector is kept for the next step of the commissioning.
Section 2 describes the design considerations of this Compton polarimeter. Section 3 reports the modifications made to 
the already existing photon beamline of a decommissioned wiggler source.
In section 4, we present the design and commissioning of the laser system and optics in the experimental hutch. 
In section 5, we report the commissioning of laser-beam interaction and the first detection of a Compton
backscattered $\gamma$ signal. Finally, we discuss the plan and perspectives in the near future.

\section{Design considerations of the Compton polarimeter}\label{sec2}

Beijing Electron-Positron Collider (BEPC\uppercase\expandafter{\romannumeral 2}) is a double-ring electron positron collider operating in the tau-charm energy range~\cite{Yu:IPAC2016cof}, with a beam energy of 1 to 2.5 \si{GeV}. It has been delivering the design luminosity of $1\times10^{33}\mathrm{cm}^{-2} \mathrm{s}^{-1}$ at 1.89 GeV in routine operation since 2023. Driven by increasing particle physics research interests at higher beam energies, it is undergoing an upgrade to achieve a threefold increase in the luminosity at 2.35 \si{GeV} and a higher beam energy coverage of up to 2.8 \si{GeV}. 
Electron beams tend to become vertically polarized in a storage ring due to the Sokolov-Ternov effect~\cite{Sokolov:S-T1963}.
According to analytical estimation, the polarization build-up time is about 70 \si{min} at 2.35 \si{GeV} for BEPC\uppercase\expandafter{\romannumeral 2} storage rings, and even shorter at higher energies.
The electron
beams are usually injected  every hour and the time-averaged polarization is expected to be larger
than $10\%$ except in the vicinity of first-order spin resonances~\cite{barber2}, according to preliminary simulations following the method in Ref.~\cite{slim}, to be reported elsewhere.
Transverse beam polarization has also some implications for BESIII
experiments~\cite{PhysRevD.110.014035}. 
Preliminary estimates suggest that a polarization degree over 10\% would be detected
by a Compton polarimeter~\cite{su_preliminary_2024}.

\begin{figure}[!ht]
\centering
\includegraphics[width=0.75\textwidth]{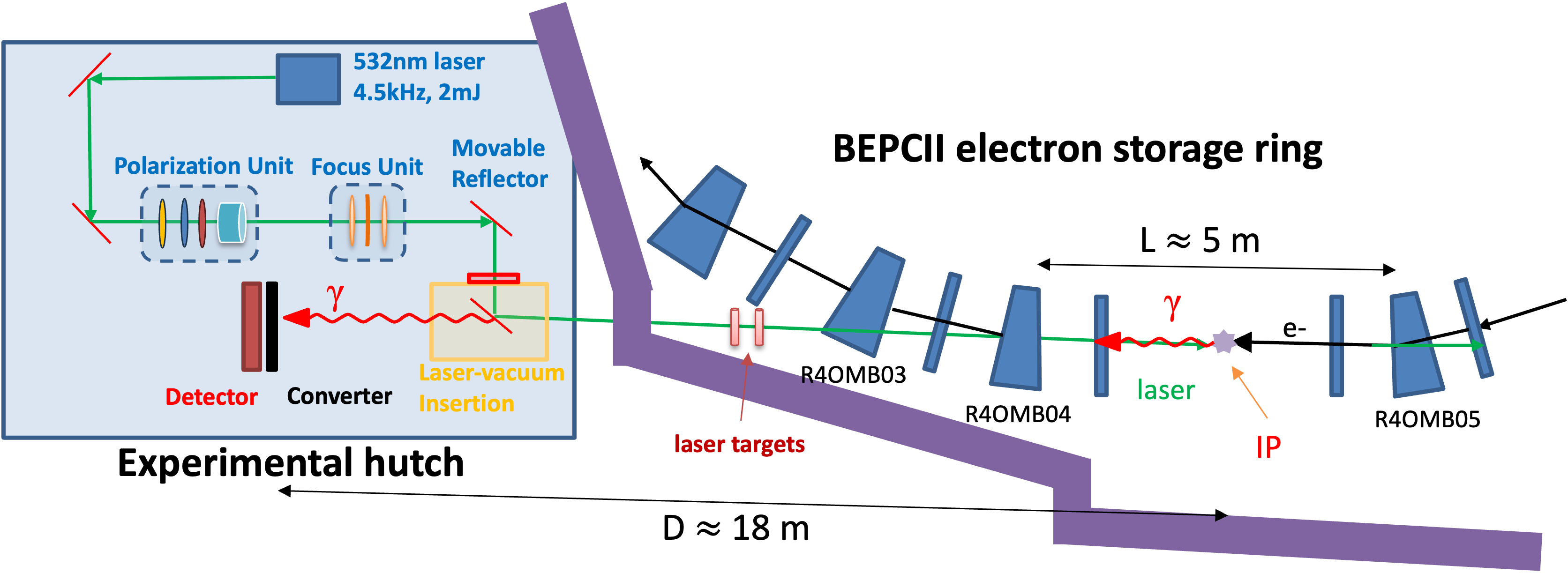}
\caption{The layout of the Compton polarimeter at the 4W2 beamline of BEPC\uppercase\expandafter{\romannumeral 2}.}
\label{fig:4W2_layout}
\end{figure}

Therefore, BEPC\uppercase\expandafter{\romannumeral 2} is a suitable test-bed to implement key technologies of Compton polarimetry.
The idea is to re-use the beamline and the experimental hutch of a dismantled wiggler beamline, 4W2.
As shown in Fig.~\ref{fig:4W2_layout},
a circularly polarized laser beam is generated in the experimental hutch and sent through the beamline
to collide with an electron bunch in the storage ring.
The back-scattered $\gamma$ photons are then transported back to the experimental hutch and detected by a detector.
This Compton polarimeter
will measure the self-polarization level in the electron storage ring, and enable resonant depolarization experiments
as well as research on radiative depolarization mechanisms and mitigation techniques,
making BEPC\uppercase\expandafter{\romannumeral 2} a suitable facility for polarized beam R\&D for future colliders.

The laser-beam interaction could take place anywhere in the straight section. For performance evaluations, we denote the center of the original wiggler section as the interaction point (IP). The corresponding distance to the detector is $D\approx18$~\si{m}.
Given the structure of the Compton cross-section given in Eq. (\ref{equ:dif_cross_section}),
a vertically segmented detector is required to measure the $P_t$.
The detector needs to cover a sensitive range of at least $\pm$5 \si{\milli \meter} and achieve a spatial resolution better than 0.1 \si{\milli \meter}.
The TaichuPix-3 silicon detector~\cite{TaichuPix-3}, 
a CMOS pixel sensor prototype for the CEPC vertex detector with 
a sensitive area of $25.7\times 15.9$ \si{\milli m}$^2$
and a pixel size of $25\times25$ \si{\micro m}$^2$, 
meets our requirements. The use of a converter made of high-Z material in front of the sensor,
that converts $\gamma$ photons to electron-positron pairs,
is likely needed to enhance the detection efficiency.
The detector and the converter are shown in red and black in Figure~\ref{fig:4W2_layout}. 

The choice of the laser wavelength needs to find a compromise between (i) a larger energy of the scattered photons by Compton scattering at lower wavelength, in order to ease its distinction from backgrounds, and (ii) the ease of operation of the laser system as a whole. Using 532 \si{\nano \meter} wavelength provides a good compromise and avoids lower wavelength that are more difficult to handle especially due to aging of optics under vacuum. This is also a wavelength that will be used for future colliders~\cite{Muchnoi_2022,CEPC_Polarimeter}.
In order to increase the rate of scattered photons, a pulsed laser beam is used. Two technologies could be used. The mode-locked laser technology provides some advantages and is expected to be used at FCC-ee~\cite{FCCee_Feasibility_Study} naturally providing high repetition rates, while gain switched lasers can provide a lower pulse repetition rate but a larger energy per pulse~\cite{Muchnoi_2022,CEPC_Polarimeter}. This second solution is employed here.
This is chosen as a compromise between (i) a significantly larger single pulse energy which will make the Compton signal more clearly visible with respect to background levels that are evenly distributed in time, and allows to measure in the future the polarization of a single bunch; and (ii) a slight decrease in the luminosity term, because Q-switched lasers typically have \si{\nano \second} pulse durations, whereas mode-locked lasers operate at \si{\pico \second} timescales.
After comprehensive comparison between different commercial products, a 532 \si{nm} pulsed laser (cnilaser, EO-532-Subnano-Hi, 2mJ),
with a repetition rate of about 4.5 \si{kHz} and a maximum pulse energy of 2 \si{mJ} has been chosen.
The laser can operate in either the internal trigger mode and the external trigger mode.
Its main parameters are summarized in Table.~\ref{tab:parameters}.
Nominally, the laser beam at the IP will be fully circularly polarized for the polarization measurement. 

The photon yield rate,
which determines the level of statistical error achieved within a given time,
is the product between the luminosity and the Compton cross-section.
For a very small crossing angle ($<$0.5 \si{\milli \radian}) as constrained by the beamline aperture,
considering that the vertical size of the electron beam is much smaller than that of the laser, the luminosity $L$ can be approximated to~\cite{Suzuki:1976xe}

\begin{align}
L &= \frac{N_e N_{\mathrm{laser}} f_{\mathrm{laser}}}{2\pi\sqrt{\sigma_{y,e}^2+\sigma_{y,\mathrm{laser}}^2}\sqrt{\sigma_{x,e}^2+\sigma_{x,\mathrm{laser}}^2+\tan{\theta}^2(\sigma_{z,e}^2+\sigma_{z,\mathrm{laser}}^2)}}\\
 & \approx \frac{N_e N_{\mathrm{laser}} f_{\mathrm{laser}}}{2\pi\sigma_{y,\mathrm{laser}}\sqrt{\sigma_{x,e}^2+\sigma_{x,\mathrm{laser}}^2}},
\end{align}
where the Piwinski contribution~\cite{Suzuki:1976xe} to the luminosity represents a $7\%$ correction to this approximation.
The number of electrons per bunch and photons per laser pulse are denoted $N_e$ and $N_{\mathrm{laser}}$, 
respectively. The collision rate is 
$f_{\mathrm{collision}} = f_{\mathrm{laser}}$
the laser repetition rate, that is much lower than the revolution frequency of BEPC\uppercase\expandafter{\romannumeral 2}. The RMS beam size of the electron and laser beams at the IP, in the horizontal ($x$) and vertical ($y$) direction are denoted $\sigma_{x/y,e/\mathrm{laser}}$. Given the electron beam parameters provided in Table~\ref{tab:parameters}, 
the luminosity $L$ can be increased by increasing the laser pulse energy, the laser repetition rate $f_{\mathrm{laser}}$ and decreasing $\sigma_{\mathrm{laser}}$ by applying laser focusing.

\begin{table}[!hbt]
   \centering
   \caption{Electron and laser beam parameters and the calculated luminosity and $\gamma$-yield rate}
   \begin{tabular}{lcc}
       \toprule
       \textbf{Parameters} & \textbf{Electron}                      & \textbf{Laser} \\
       \midrule
        Energy(eV)      & $2.35\times 10^9$         & 2.33   \\ 
        e- bunch population &  $3.7\times10^{10}$  &  -  \\
        Pulse energy(mJ)  &  -  &  2  \\
        Repetition rate(kHz)  & $1.26\times 10^3$   & 4.508   \\
        Pulse length(ps)  &   50       & 1277   \\ 
        Rms emittances(nm) & 157/0.53     & -     \\
        Beta function(m) &  5.66/5.31 & - \\
        Rms energy spread & $6.35\times10^{-4}$ & -\\
        Dispersion function(m) & 1.32/0	 & - \\
        Rms beam size(mm)  & 1.3/0.05   &  0.5/0.5     \\ 
        \midrule
        Horizontal crossing angle(mrad) & 0.5 \\
        Luminosity (barn$^{-1}$s$^{-1}$) & $2.1\times10^7$  \\
        $\gamma$-yield per crossing & 2842 \\
        $\gamma$-yield rate(MHz) &   12.8  \\
        Statistical uncertainty  &  ~0.9\% @ 16 \si{\second} \\
       \bottomrule
   \end{tabular}
   \label{tab:parameters}
\end{table}

Table~\ref{tab:parameters} presents the parameters of the electron and laser beams, 
along with the calculated luminosity and the $\gamma$-yield rate.
As will be described later, simulations of the laser focusing unit suggested that
an RMS laser beam size as small as 0.5 \si{\milli \meter} is achievable at the IP, further reduction
is otherwise constrained by the aperture limitation of the beamline chamber.
Monte Carlo simulations showed that by measuring $1\times10^8$ Compton scattering events of each laser helicity, a statistical uncertainty of about 0.9\% could be obtained~\cite{su_preliminary_2024}. 
This corresponds to a data taking of about 16~\si{\second}, with the current experimental setup, where the repetition rate and average power of the laser has been updated relative to those parameters assumed in Ref.~\cite{su_preliminary_2024}.

\section{Modification of the photon beamline}\label{sec3}

 The layout of the Compton polarimeter requires the transmission of both the laser beam and back-scattered $\gamma$ photons through the same photon beamline of the dismantled wiggler.
 Moreover, the synchrotron radiation heat load in the beamline
 is substantially reduced due to the removal of the wiggler magnet, and now only synchrotron
 radiation from the exit of the upstream bending magnet and the entrance of the downstream bending magnet
 can reach the experimental hutch.
 The existing beamline has been optimized for the transmission of high-flux X-rays at \si{\kilo eV} energies.
 In order to minimize implementation cost and efforts, we decided to reuse most of the beamline elements,
 in particular the existing vacuum system, radiation safety measures and the experiment protection system.

\begin{figure}[!htb]
    \centering
    \includegraphics[width=1.\columnwidth]{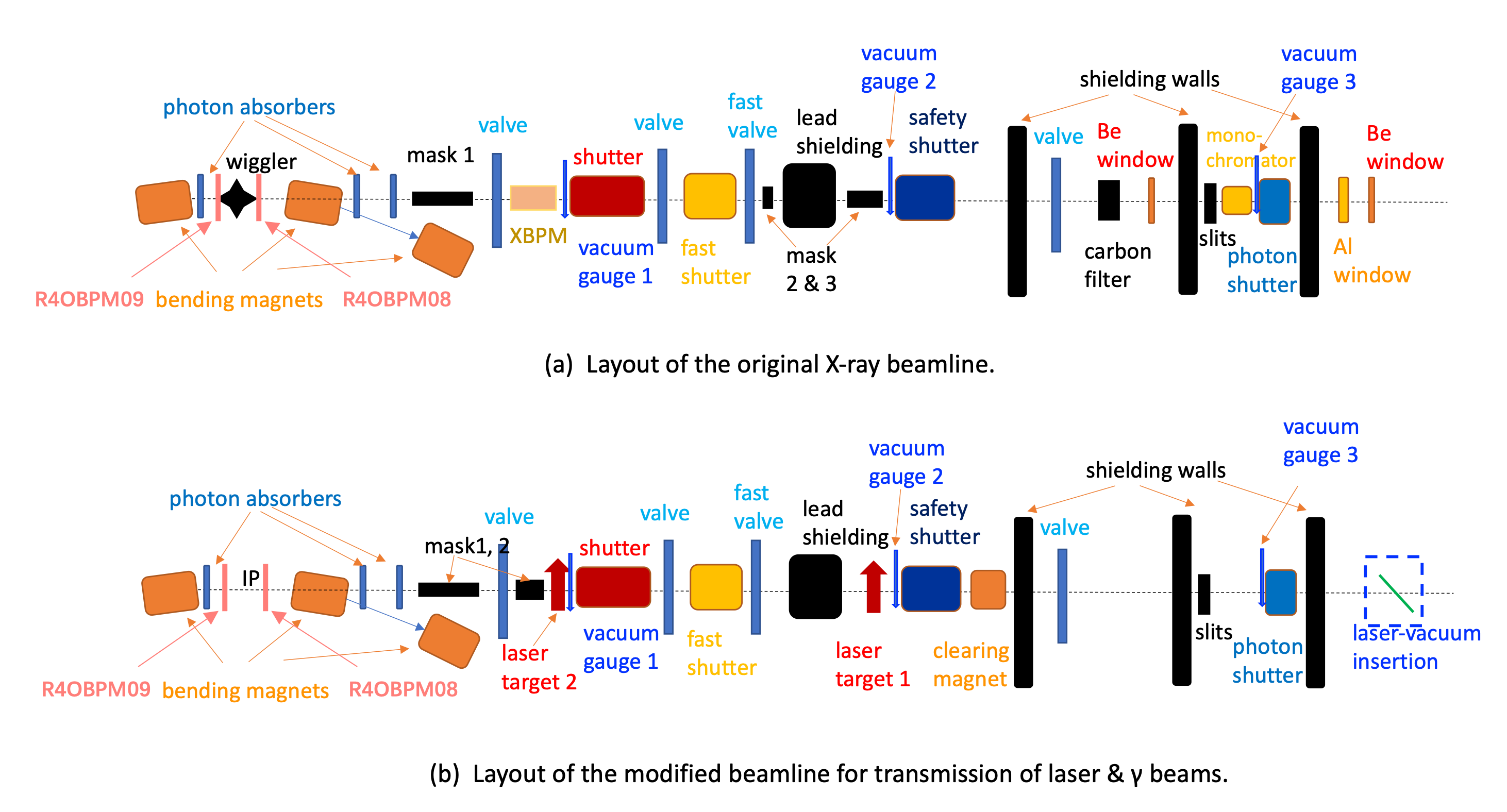}
    \caption{The component layout and modifications of the photon beamline.}
    \label{fig:beamline_layout}
  \end{figure}

 The component layout and modifications of the photon beamline are shown in Fig.~\ref{fig:beamline_layout}. 
 Beamline components like Beryllium and Aluminum windows and the carbon filter that are opaque to green light laser beam were identified 
 and replaced by more suitable vacuum chambers, the monochromator is no longer in use.
 Two water-cooled masks, implemented to absorb the X-ray heat load, were now considered to be limiting the acceptance angle of the $\gamma$ photons and also removed, 
 a new water-cooled mask (mask 2 in Fig.~\ref{fig:beamline_layout}(b)) was installed in the place of the XBPM 
 to protect the parts of the
 safety shutter with no water cooling from radiation heat load. 
 
\begin{figure}[!htb]
    \centering
    \includegraphics[width=0.8\columnwidth]{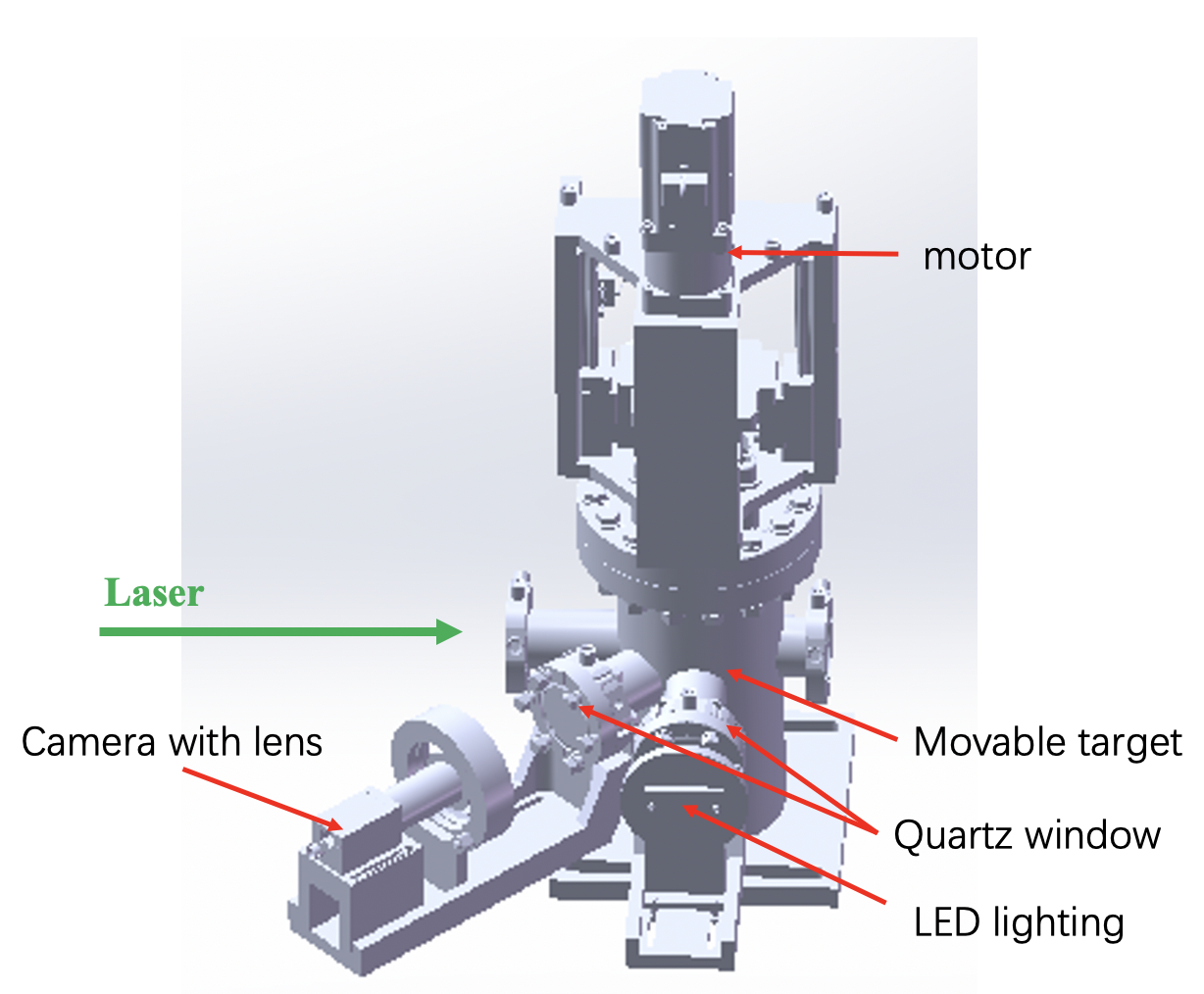}
    \caption{A schematic illustration of one laser alignment target installed in the front end.}
    \label{fig:target}
\end{figure}

Two laser alignment targets were implemented into the beamline for diagnostics of the position and size of the laser beam, using the space of an XBPM and a fixed aperture. 
As shown in Fig.~\ref{fig:target}, each laser alignment target has a vertically movable ceramic target inside the chamber.
When the target is positioned in the path of the laser, the diffusely reflected light at approximately 45 degrees is captured by an imaging system connected to a CMOS camera~\cite{su_preliminary_2024}.
The image of the laser spot on the camera is then fitted with a Gaussian distribution to extract the center position and spot size of the laser.

These modifications were carried out during the long shutdown of 2024, followed by an experiment 
that confirms the laser beam could be transported from the hutch to the planned IP, ensuring there is no obstacle in the beamline, and establishing a reference trajectory for the laser with recorded positions
on the two laser alignment targets. BEPC\uppercase\expandafter{\romannumeral 2} resumed beam operation in May 2025,
after several weeks of vacuum conditioning,
the readings of the vacuum gauges 1, 2 in the front end region were typically $3\times10^{-9}$ Torr while the reading of the vacuum gauge 3 near the experimental hutch
was typically $2\times10^{-8}$ Torr, respectively, indicating that with these modifications a reasonable well vacuum level is still attainable
in the beamline. 

\section{Laser optics tuning \label{sec4}}

\begin{figure}[!ht]
\centering
\includegraphics[width=1 \textwidth]{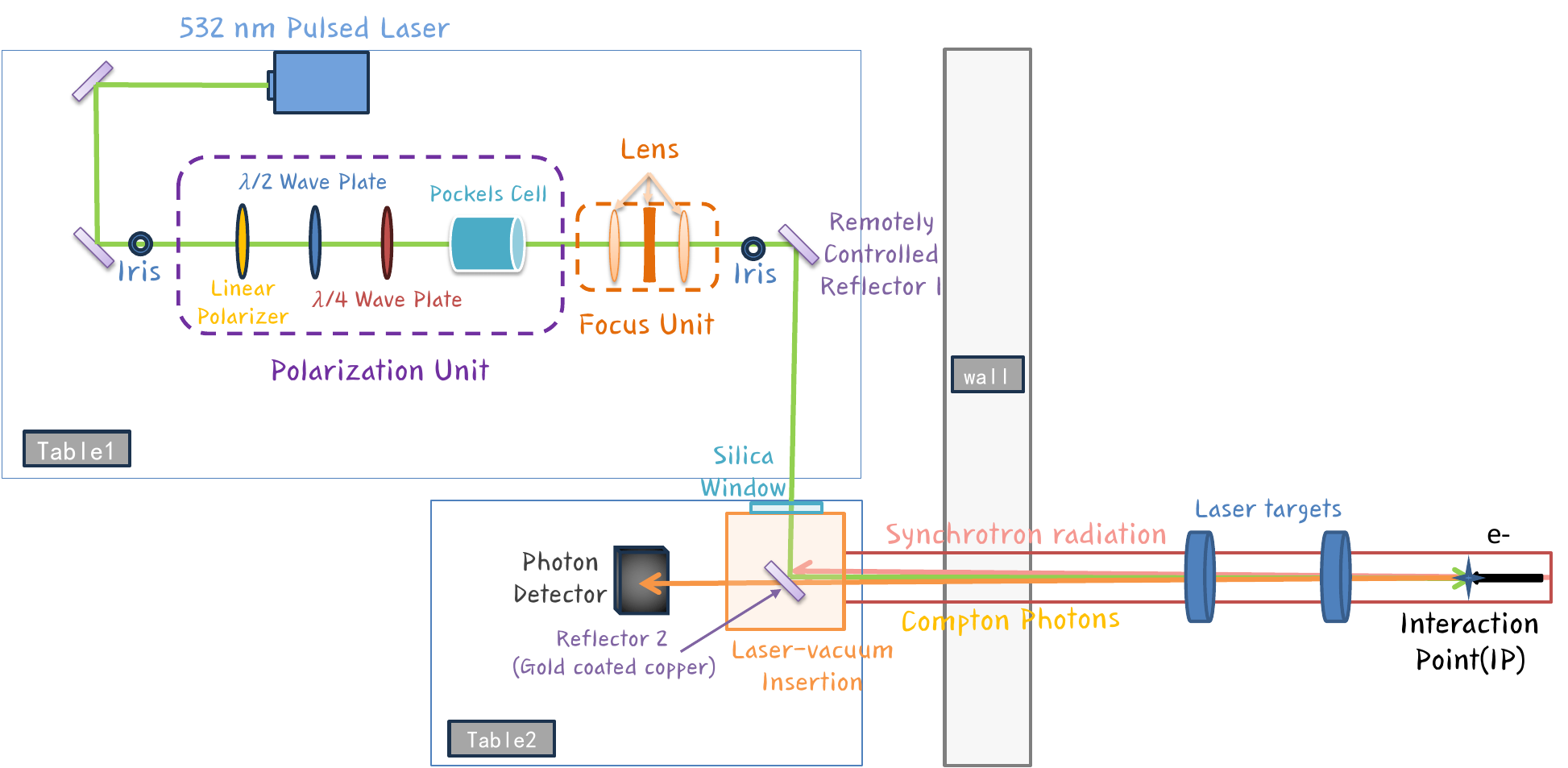}
\caption{
The components of the Compton polarimeter at the 4W2 experimental hutch of BEPC\uppercase\expandafter{\romannumeral 2}.}\label{fig:hutch}
\end{figure}

Following the principle of the Compton polarimeter as detailed in Section \ref{sec2},
the laser optics system needs to transport a focused, circular polarized laser beam 
with switchable laser helicity, to collide with the electron beam at the IP.
The modular design of the laser optics system is shown in Fig.~\ref{fig:hutch},
its design, implementation and tuning will be
elaborated in this section.

\subsection{polarization unit\label{subsec1}}

The polarization unit is designed to ensure a specific circular polarization state at the IP. 
The inherent polarization ratio of the laser beam is $T_p:T_s=149:1$. After passing through the first element of the polarization unit, a linear polarizer, the polarization ratio is enhanced to $T_p:T_s>5000:1$.
The conventional scheme to generate circularly polarized light utilizes a quarter-wave plate placed at 45 degrees relative to the
linear polarizer. 
But we found that as the laser is transmitted toward the IP, the subsequent elements alter the circular polarization significantly. 
To this end, we employ both a $\lambda/2$ wave plate and a $\lambda/4$ wave plate
for compensating the effect of all preceding optical elements to maximize the attainable degree of circular polarization.
Next, the Pockels cell can be set to either 0 voltage or 1/2 wavelength voltage, to
switch the helicity of the circular polarization.
The commercial instrument, PolSNAP Stokes Polarimeters from Hinds Instruments, was used for measuring the beam polarization at the place between reflector 2 and the beamline elements, before the vacuum chambers were connected. Using a 532~\si{\nano \meter} CW laser, we have demonstrated the procedure of laser polarization tuning, with switching of left- and right-handed helicity at a frequency of 0.1 Hz, and the reading of the circular polarization better than $99\%$. However, as will be discussed in Section 4.3, the laser wavefront distortion must be mitigated before a laser beam with a reliable circular polarization can be delivered. In future development, we will also implement a real-time laser polarization measurement module.

\subsection{Alignment and focusing units}\label{subsec2}

Online adjustment of the laser position at the IP is achieved by reflector 1, which can be rotated along both pitch and yaw axes remotely controlled by stepper motors.
The RMS horizontal and vertical sizes of the electron beam are 
1.3~\si{\milli \meter} and 0.05~\si{\milli \meter} 
at the IP, respectively, smaller than those of the laser beam.
At the moment, the resolution of the rotation angle adjustment is 
16.4 \si{\micro \radian} along each axis,
corresponding to 0.216 \si{\milli \meter} per step at the IP. This is acceptable for the coarse
tuning of the laser-electron collision, but insufficient for fine optimization
of the luminosity.
Therefore,
further improvements of the adjustment step size and also the bidirectional repeatability are planned.

\begin{figure}[!htb]
    \centering
    \includegraphics[width=0.8\columnwidth]{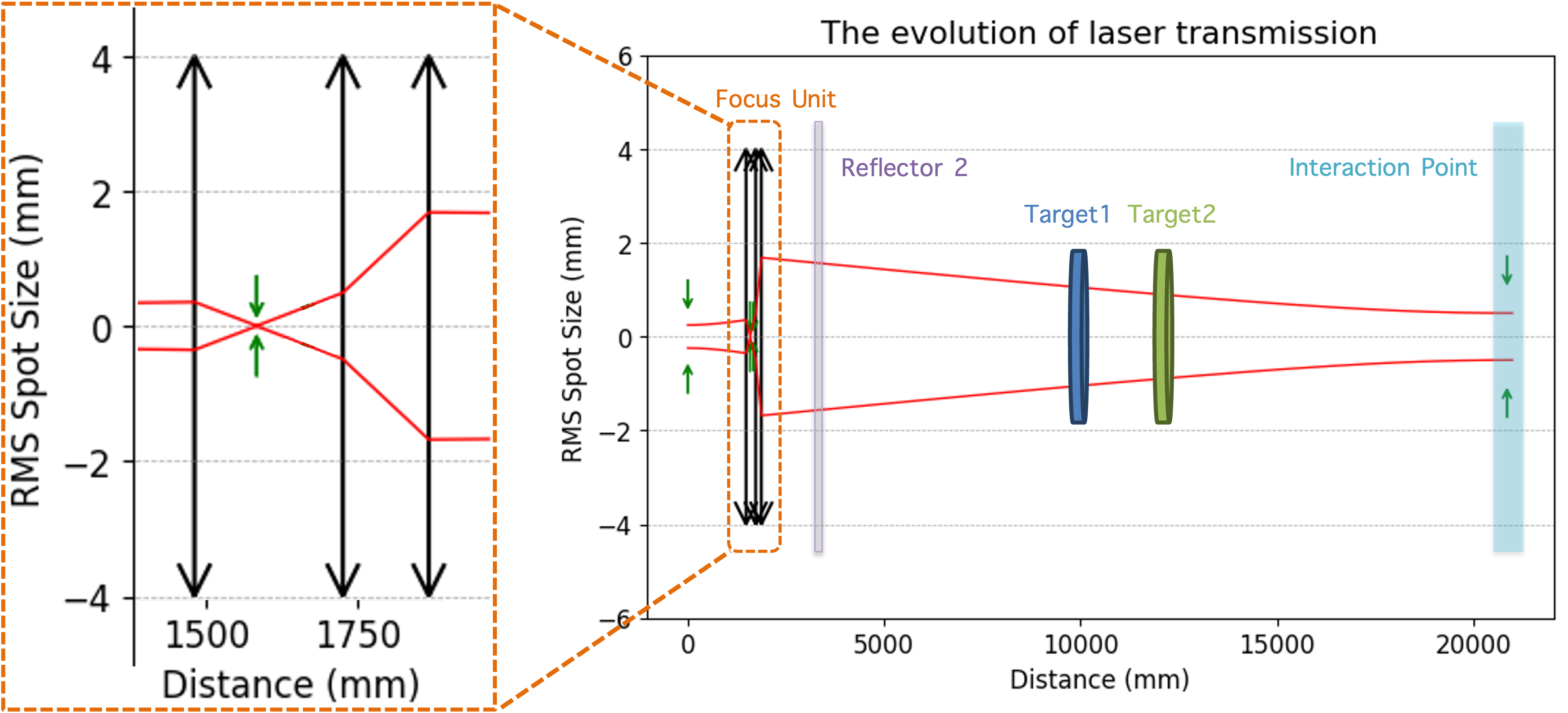}
    \caption{The evolution of the RMS spot size of the laser from the source to IP, the region of the focusing unit is highlighted.
    The double arrows point to the locations of the laser beam waist. 
    }
    \label{fig:laser&lenses}
\end{figure}

To focus the laser beam at the IP, about 20~\si{m} away from the laser, a focusing unit with a long
effective focal length is required. 
To this end, we chose a set up of 
three lenses with focal lengths of 100 mm, -100 mm, and 200 mm 
respectively, and the relative distances between these lenses can be manually adjusted. 
Based on calculations using Gaussian beam optics, 
the focusing unit can be optimized by simulating the evolution of the laser beam size within the beamline.
For a laser beam with an initial spot size of $\sigma_{\mathrm{laser0}}=0.25$ \si{\milli \meter} and a full divergence angle 
of $1.35$~\si{\milli rad}, 
after optimizing the focusing unit, 
the evolution of the laser beam size is shown in Fig.~\ref{fig:laser&lenses},
with the laser beam focused to $\sigma_{\mathrm{laser}}=0.5$ \si{mm} at the IP.

\subsection{Laser-to-vacuum insertion\label{subsec3}}
Two reflectors are used to guide the laser beam to the IP. Reflector 2 is placed between the IP and the detector, in a very constrained space. As a consequence, it is in the path of the synchrotron radiation. 
Moreover, there must be a window to seal the vacuum, which (i) demonstrates excellent sealing capabilities, can withstand radiation without compromising the vacuum, (ii) has high transmittance for the 532~\si{\nano \meter} laser wavelength, allowing efficient laser beam passage while minimizing energy deposition and temperature rise and (iii) is non-birefringent and remains stable optical performance when exposed to synchrotron radiation.
This laser-to-vacuum insertion
is one important aspect of the Compton polarimeter.

\begin{figure}[!htb]
    \centering
    \includegraphics[width=0.8\columnwidth]{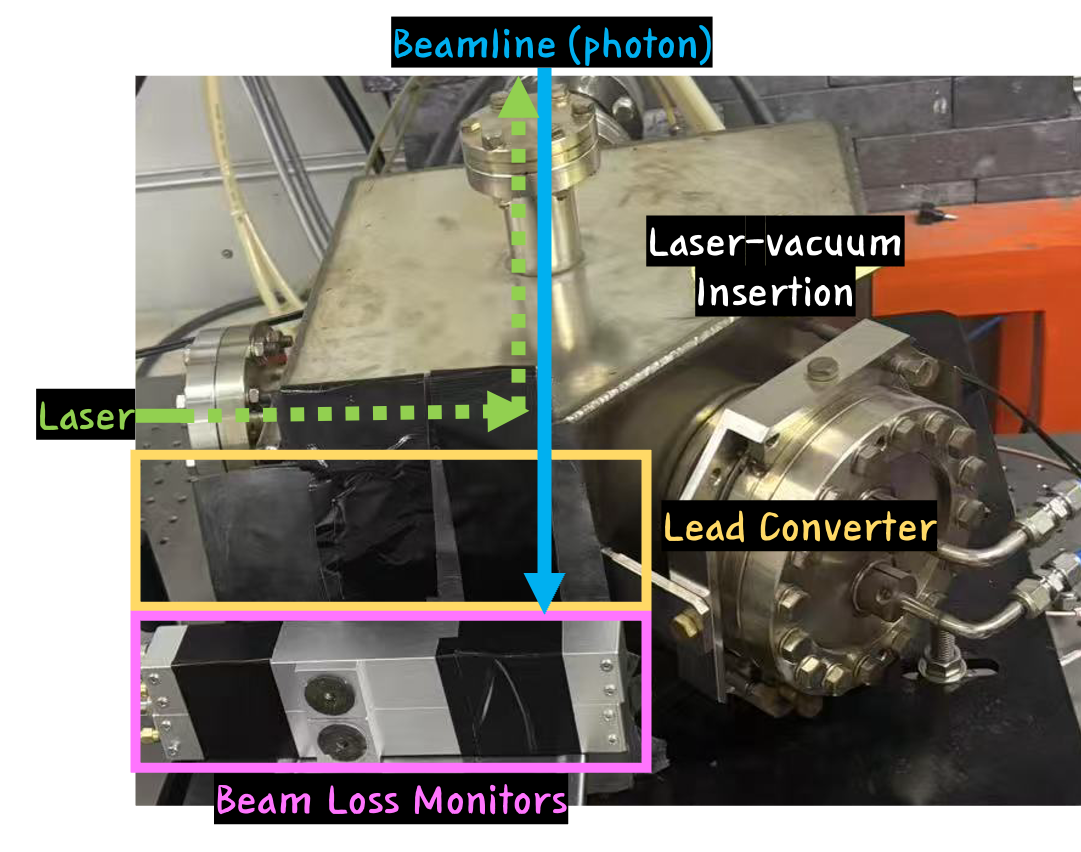}
    \caption{ The photon of the laser-to-vacuum insertion and nearby region.
    The laser beam (green dashed arrow) is deflected by reflector 2 inside the laser-to-vacuum insertion towards IP.
    The beam background and signal (blue arrow) pass through the laser-to-vacuum insertion and a lead converter (yellow box),
    and are detected by two BLM detectors (violet box) placed one above the other.}
    \label{fig:BLM}
\end{figure}

To this end, we borrowed a used laser-to-vacuum insertion~\cite{abakumovaSystemDeliveryIR2015} from the BEPC\uppercase\expandafter{\romannumeral 2} beam energy measurement system~\cite{xhmo:biienergy2010,jyzhang:biienergy2016}, which
utilized laser Compton backscattering to measure the electron beam energy and rms energy spread. This laser-to-vacuum insertion encloses a water-cooled, gold-coated copper mirror (reflector 2)
by a stainless steel vacuum chamber.
As shown in Fig.~\ref{fig:BLM}, after the laser-to-vacuum insertion was connected to the vacuum chamber of the beamline, the laser beam entered the vacuum chamber through a fused silica window, and was deflected by reflector 2 towards the IP.

\begin{figure}[!ht]
    \centering
     \subfigure[Laser target 1]{
         \begin{minipage}[ht]{1 \textwidth}
             \centering
             \includegraphics[width=0.8 \textwidth]{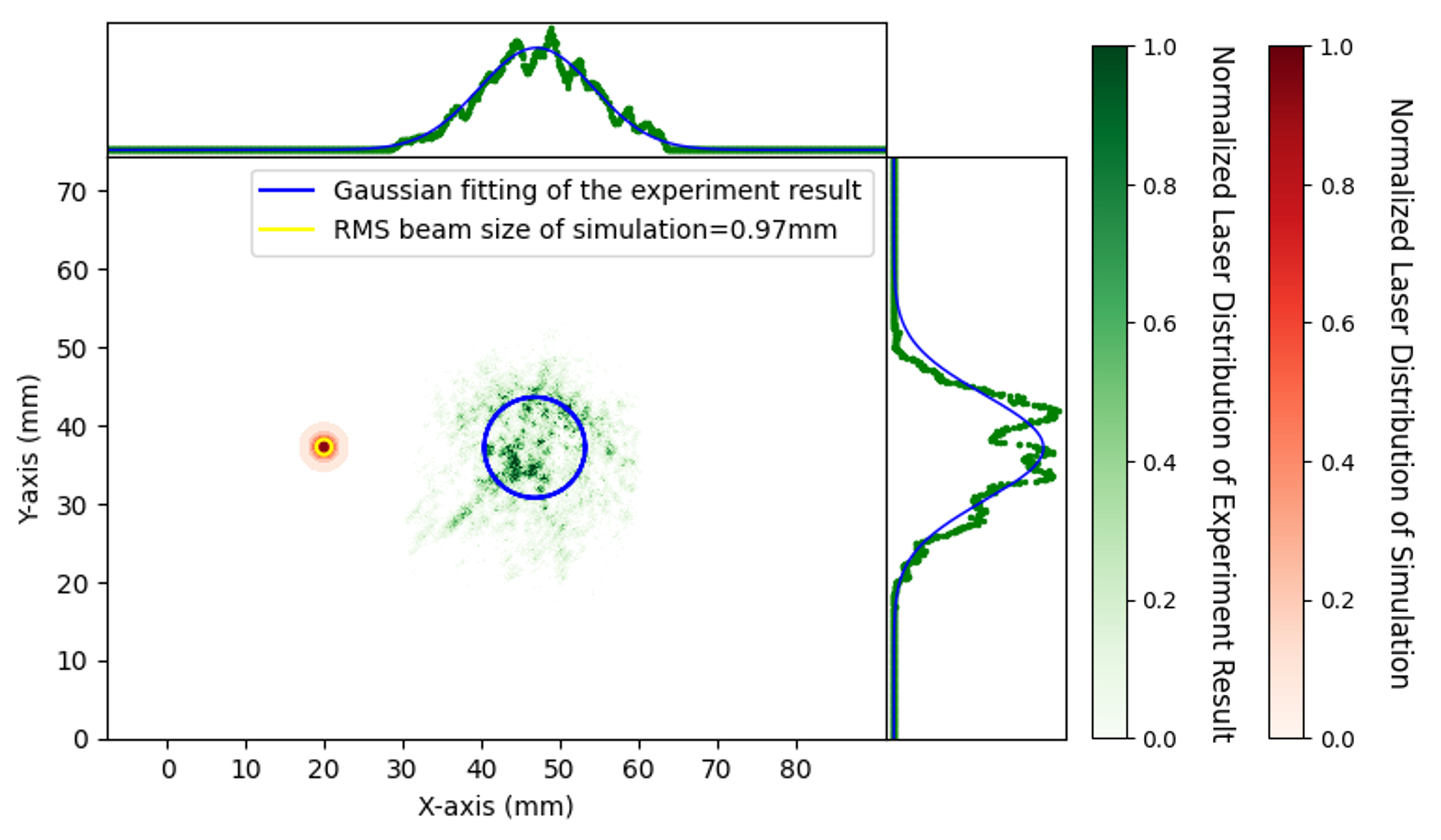}
         \end{minipage}%
     }
     \subfigure[Laser target 2]{
         \begin{minipage}[ht]{1 \textwidth}
             \centering
             \includegraphics[width=0.8 \textwidth]{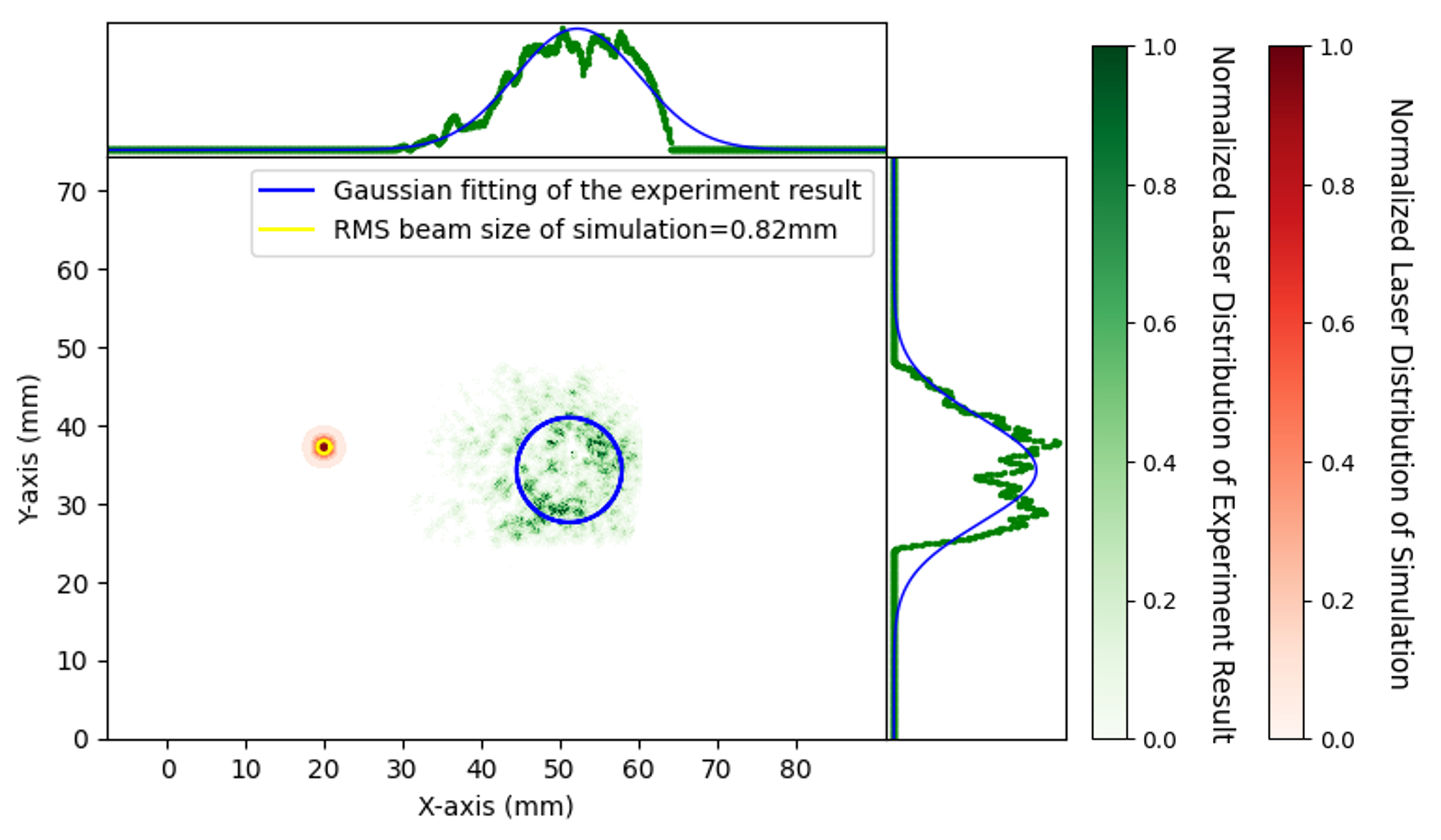}
         \end{minipage}%
     }%
   \caption{
   The laser beam transverse distribution on laser target 1 (upper plot) and laser target 2 (lower plot). The green dots show the normalized laser beam distribution
   from the measurement and the blue ellipse shows the fitted RMS beam sizes.
   In contrast, the simulated normalized laser beam distribution (in red) and the fitted
   RMS beam sizes (yellow circle) are plotted with a horizontal offset of 
   about -25 \si{mm}.
   }
    \label{fig:spots on target}
\end{figure}

When the focusing unit was set up accordingly,
however, the measurements of the beam sizes on the laser target, 
as shown in Fig.~\ref{fig:spots on target},
did not agree with the simulations in Section \ref{subsec2}.
After adjusting lens distances to minimize the beam spot size at the laser targets,
the laser spots on both targets were still much larger than the simulation results.
In particular, it appears that the laser beam exhibits severe wavefront distortion.
The laser targets are known to behave properly, as tested during early commissioning when the tube was not evacuated and the laser vacuum insertion was not installed. Moreover, we could check that significant distortion of the laser profile was also observed right after the laser-vacuum insertion, by placing a mirror to redirect the laser beam, which shows that this is not likely related to internal reflections in the beam pipe.
Indeed it has been designed for operation with a CO$_2$ laser, i.e. with a wavelength 20 times larger than what is implemented in the experiment described here. As a consequence, sensitivity to surface roughness is increased. Moreover, this laser-to-vacuum insertion was
exposed to synchrotron radiation during years of operation and then stored in the ambient atmosphere for one year, possibly leading to degradation of the mirror surface. One possible mitigation is to replace the laser-to-vacuum insertion with a water-cooled diamond window and an out-of-vacuum reflector. 

\section{Beam commissioning \label{sec5}}

During the accelerator operation between May and July, 2025, 
BEPC\uppercase\expandafter{\romannumeral 2} storage rings were operated at a beam energy of 1.84~\si{GeV}.
The electron ring was refilled normally every 1 hour, between
adjacent injections the circulating beam current decays from about 600 \si{\milli A} to 400 \si{\milli A}.
As the RF frequency is 499.8~\si{MHz}, there are in total 396 RF buckets along the 237.53 \si{meter} long electron storage ring 
with a bucket spacing of about 2 \si{\nano \second}.
Typically one bunch train of around 100 bunches are filled for electron-positron collisions with a 6 \si{\nano \second} bunch spacing, i.e., 1.8 m distance between adjacent bunches, with a large gap between the head and tail of the bunch train for ion clearing. 

According to simulations and previous measurements at this experiment hutch, the beam background noise contains spurious electrons with energies up to the beam energy,
synchrotron radiation with a critical energy of 1.5~\si{keV}, and $\gamma$ photons from beam-gas bremsstrahlung with energies also up to the beam energy.
In order to measure the beam background and signal in real time, we employed beam loss monitors (BLMs)~\cite{yu_BLM2024}
as the detector
system.
A BLM
contains 
a 
plastic scintillator
coupled with 
a 
Hamamatsu H10721-110
photomultiplier tube (PMT).
The scintillator is a EJ-200 rod
with 
a diameter of 22~\si{\milli\meter} and a length of 100~\si{\milli\meter}
wrapped into a high
reflectivity 
aluminium foil.
As shown in Fig.~\ref{fig:BLM},
two BLMs were positioned one above the other,
placed behind a 3-mm-thick lead
converter
facing the beamline chamber.
Such a detector is
sensitive to spurious
electrons, synchrotron radiation and $\gamma$ photons from
both bremsstrahlung of electrons with residual gas, and the Compton scattering.
The size of the total
sensitive area of the two BLMs is approximately $4.5 \times 10$ \si{cm}$^2$, 
such layout of the detectors was chosen to cover
the whole range of the vacuum pipe.
The time resolution
of BLMs is on the order of 10 \si{\nano \second}.
The output signals of both BLMs are connected to two different channels
of an oscilloscope, and the event rates are recorded. The lead 
converts $\gamma$ photons to electron-positron
pairs and enhances the detection efficiency for $\gamma$ photons,
it also substantially reduces the flux of synchrotron radiation photons.
When the lead converter was inserted, a roughly three time increase of the event rate was observed, suggesting that the bremsstrahlung $\gamma$ photons are
the predominant background contributor as detected by BLMs. Note that both the $\gamma$ photons from the Compton scattering and from the bremsstrahlung follow
closely the tangential direction of the electron beam trajectory at the straight section, it is essential to align the electron beam, the mechanical apertures in the beamline and the detector.

Firstly, we performed a coarse alignment of the detector with the beamline by adjusting the positions of the BLMs in both the horizontal and vertical directions
to maximize the detector event rates.
Additionally, we noted that there is
no obvious difference in the
signal amplitude and the event rate 
when we switch on and off the laser, indicating that these signals are
predominantly
background noise. 

\begin{figure}[!htb]
    \centering
    \includegraphics[width=0.8\columnwidth]{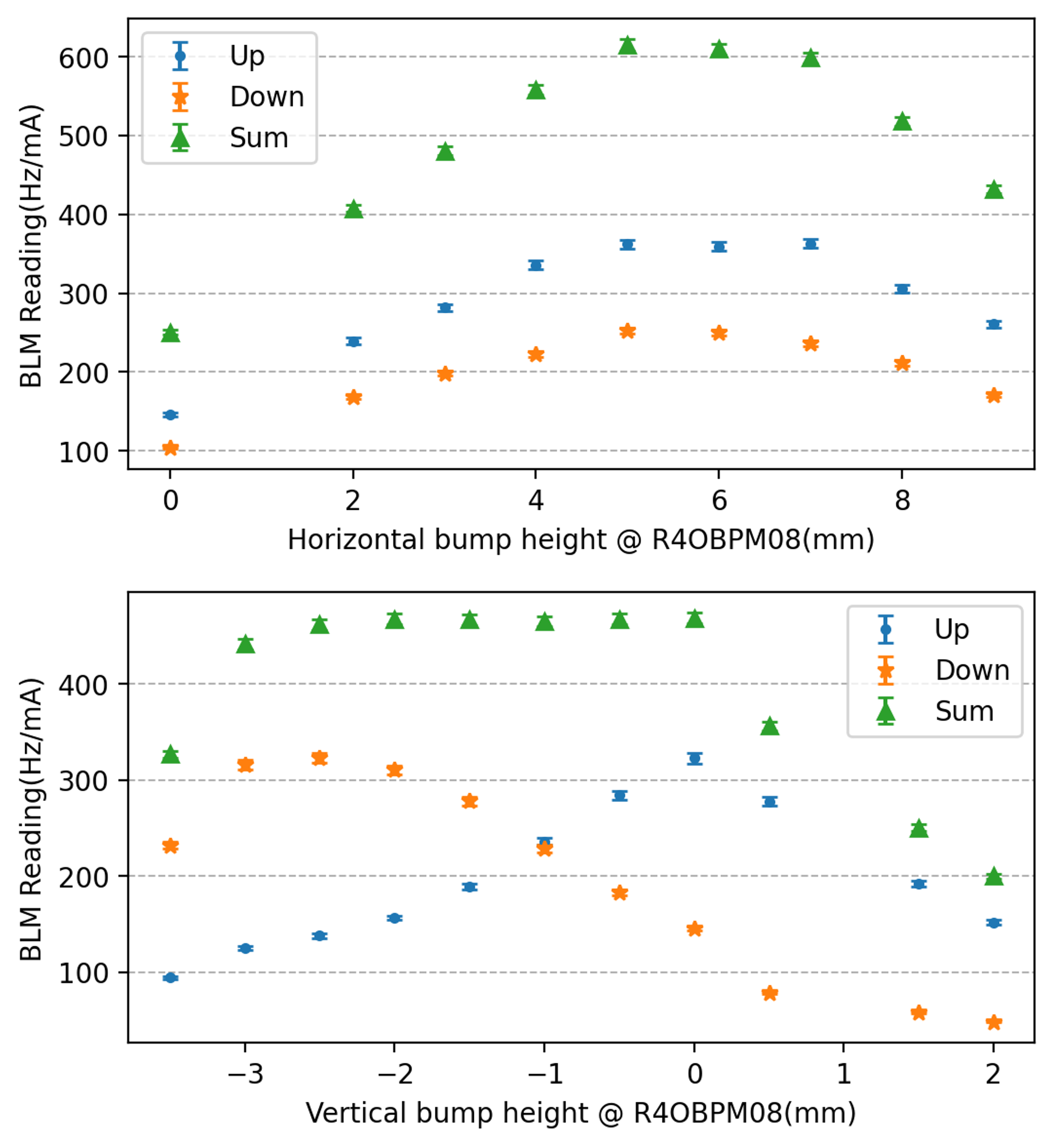}
    \caption{Without laser interaction, the reading of upper BLM, lower BLM and sum of both vary when we adjust the center position of electron beam horizontally and vertically based on the local bump method. 
    It is noteworthy that the two beam loss monitors are vertically positioned and aligned horizontally. As a result, during vertical scanning, the signal rates peak sequentially, while during horizontal scanning, they reach their peak simultaneously.}
    \label{fig:BLM_reading}
\end{figure}

Next, we aligned the electron beam trajectory at the straight section, to maximize the beam background signal rate detected by BLMs.
Several correctors were
employed to set up horizontal and vertical closed orbit bumps.
The position at the beam position monitor R4OBPM08, which is near the IP, was scanned, and the detected event rate was recorded. 
As shown in Fig.~\ref{fig:BLM_reading}, during the horizontal scanning, the event rates of both BLMs reach their peaks simultaneously, while during the vertical scanning, the signal rates reached their peaks sequentially, which is consistent with the fact that both BLMs are placed one above the other. 
We thus obtained the reference trajectory of the electron beam when it was roughly aligned with the 4W2 beamline and the BLMs. 

\begin{figure}[!htb]
    \centering
    \includegraphics[width=1 \columnwidth]{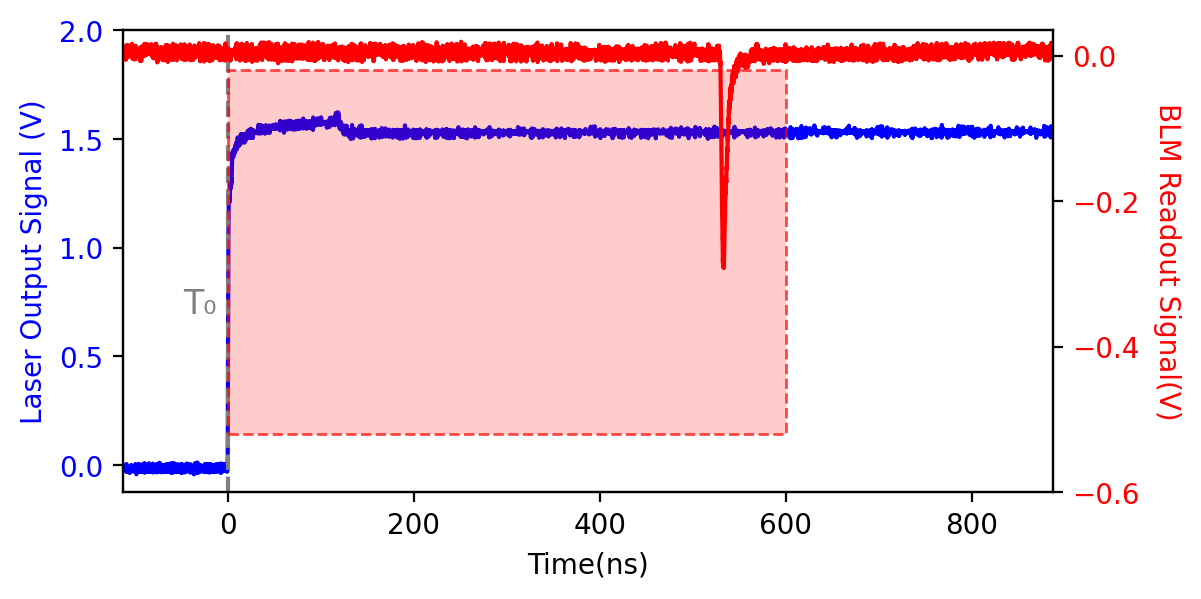}
    \caption{A selected event observed on the oscilloscope. The rise edge (denoted as $T_0$) of the square waveform from the laser output (blue curve) is used as the trigger signal. 
    A pulse signal from the upper BLM (red curve) is selected, with its voltage between [-0.52 V, -0.02 V] and rise time between [0 ns, 600 ns] after the trigger time $T_0$.}
    \label{fig:trigger600ns}
\end{figure}

Following this, tuning towards attaining laser-electron collisions was launched. 
The laser was operated in the internal trigger mode with a fixed repetition rate of 4.5 \si{\kilo Hz}, asynchronous with the ring RF frequency.
On average, each laser pulse will coincide with more than 4 electron bunches in the 5-meter-long straight section.
Different from the beam background signals
which are considered to be independent of the laser pulses, the detection of the Compton backscattered $\gamma$ photons by BLMs must follow 
the output of the laser pulses by a 
time delay of a few hundred nanoseconds, taking into account of the path length of the laser and $\gamma$ photons,
the delays due to the laser, the detector and cables. 
Using this causality relation,
the oscilloscope was set up to capture and save
the data for the events detected by the upper BLM.
A typical selected event is
illustrated in Fig.~\ref{fig:trigger600ns}, 
the rise edge of the square waveform from the diagnostic port of the laser 
is denoted as $T_0$, the selected events must satisfy the following two conditions: (i) the signal voltage $V$
exceeds a specified voltage threshold $V_{th}$,
to suppress the background unrelated to the beam, $V_{th}$ was set to 20 mV empirically; (ii) 
the rise time $t$ of the signal pulse is within a range
[$T_0$, $T_0+t_{max}$], $t_{max}$ was set to 600 ns in the initial commissioning of laser-electron collision. 

\begin{figure}[!htb]
    \centering
    \includegraphics[width=1 \columnwidth]{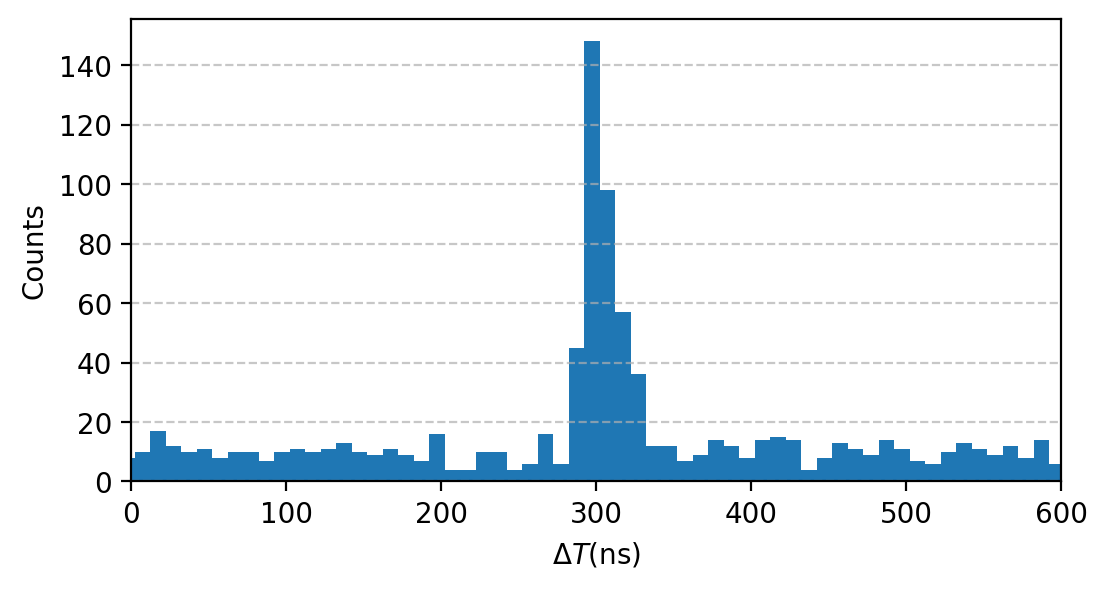}
    \caption{Histogram of the relative time $\Delta{T}=t-T_0$ of the peak of signal pulses among 1000 selected data sets.}
    \label{fig:causality}
\end{figure}

The laser trajectory was scanned in two orthogonal directions, by adjusting the yaw and pitch angles of the movable reflector 1.
Laser target 1 was used for rough positioning of the laser, as the centers of target 1, target 2 and the IP have been roughly aligned before. 
As mentioned in Sec.~\ref{subsec2}, each step of the angular adjustment of reflector 1 corresponds to a position change of 
0.216~\si{\milli \meter} at the IP. 
Then a 2D grid scan of the two angles of reflector 1 was launched to maximize the detected event rate, identifying 
a position with an event rate over ten times higher than the background level, reflector 1 was secured at this position. 
We stored the waveform of the BLM reading of each triggered event, and acquired a total of 1000 sets of waveforms. 
Then we calculated the relative time of the pulse peak
$\Delta{T}=t-T_0$
in each event, and plotted its histogram among these 1000 data sets, as shown in Fig.~\ref{fig:causality}.
This distribution indicates a clear dependence of the count of the detected signal
on the delay of the signal detection relative to the laser emission, peaked around 300~\si{ns}, which exhibits a correlation between the laser and the detected signal.

\begin{figure}[!htb]
    \centering
    \includegraphics[width=1 \columnwidth]{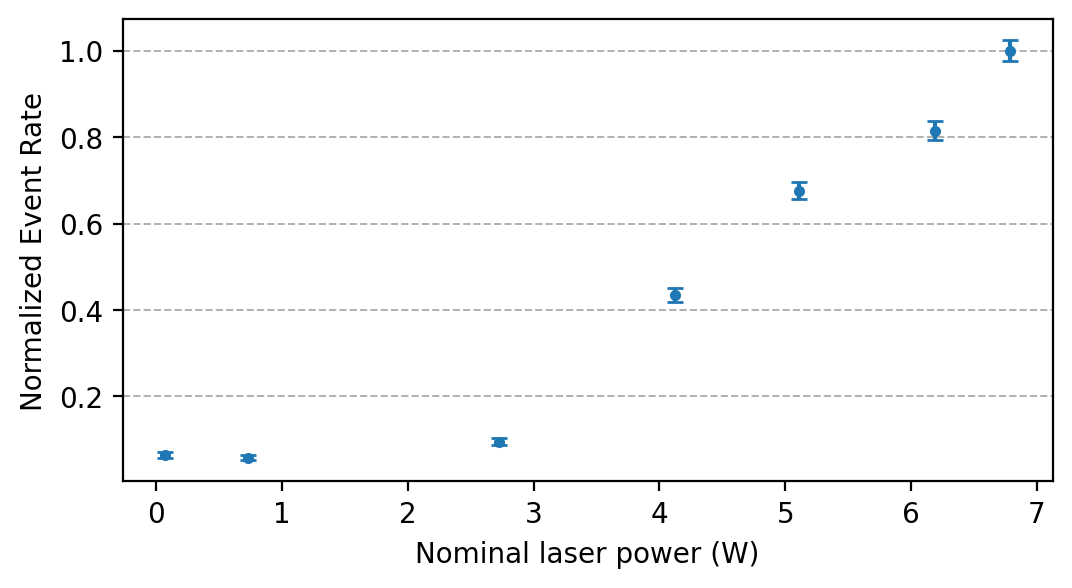}
    \caption{The normalized event rates for different settings of the nominal laser power.}
    \label{fig:Event rate vs laser power}
\end{figure}

Furthermore, we scanned the current of the laser source and recorded the event rate.
Note that we were not capable of conducting real-time measurement of the laser power, and some of the event rate data was not taken when the laser became stable
at the laser current setting. 
Instead, we calculate the nominal laser power as an approximation, using the offline calibrated current-power table.
Fig.~\ref{fig:Event rate vs laser power} shows the dependence of the event rate, normalized to its maximum value, on the nominal laser power. 
At a higher nominal laser power, the general trend is that the event rate increases approximately linearly with the increase in the nominal laser power.
These results confirmed the successful detection of Compton back-scattered signals.

In addition, the BEPC\uppercase\expandafter{\romannumeral 2} timing system was configured to send external triggers to 
the laser at a frequency of 4.508 kHz, $1/280$ of the revolution frequency. 
To this end, we injected a non-colliding pilot bunch in the abort gap.
By adjusting the trigger delay, 
the precise synchronization between the laser pulses and this pilot bunch was achieved.
Using the same detection method as described before,
the back-scattered $\gamma$ signals from the laser-electron collisions were also observed. This paves the way for
measuring the beam polarization of arbitrarily selected electron bunches in the storage ring.

\section{Outlook}\label{sec7}
During the development of the Compton polarimeter, 
we have identified several limitations in the existing system, and improvements are foreseen.
In addition, we have also planned the next stage beamline modification, as well as the procedures of upcoming beam commissioning.

The laser-to-vacuum insertion is essential for the operation of the Compton polarimeter,
a new solution is thus necessary to mitigate the laser beam wavefront distortion
which persists in the existing system and leads to deterioration of
the collision rates. One approach is to use a water-cooled 
diamond window to seal the vacuum
and absorb the soft X-ray in the synchrotron radiation,
and an out-of-vacuum reflector to deflect the laser beam, which is easier to handle and replace.
Preliminary tests with such a layout show great improvements in the quality of laser focusing,
while more beam tests are necessary to evaluate its influence to
beam polarization measurements and robustness in long-term operation.

The tuning and optimization of the laser-electron collision calls for improvements in the
laser alignment and focusing system. 
A finer step size and better bidirectional repeatability of the laser angular tuning is important for the fine tuning of the overlapping between the laser and electron beams. 
A remotely controllable precise tuning of the distances between the lenses is essential for
continuous optimization of the laser focusing at the IP. 
It is also beneficial to have
additional diagnostics of the spot size and the polarization state of the laser beam 
on the other end of the IP, which requires some nontrivial
modification work in the storage ring and more studies are needed.

 As BEPC\uppercase\expandafter{\romannumeral 2} will start operation at a beam energy of 2.35 GeV in early 2026, two new photon absorbers
 with slits for photon beam transmission,
 that can withstand a higher synchrotron radiation heat load must be installed. They will be located in the electron storage ring
 between the laser-electron IP and the beamline, in addition to a new anti-chamber that
 accommodates one of these photon absorber.
 In addition, a permanent dipole magnet will also be installed
 in the beamline, to prevent spurious electrons from reaching the experiment hutch.

In the upcoming run of BEPC\uppercase\expandafter{\romannumeral 2}, we will pursue beam
commissioning of the TaichuPix-3 detector. 
A method to synchronize the laser pulse and the TaichuPix-3 detector is necessary for observation of
the laser-electron collision signals. Then a proper data taking procedure needs to be
established to integrate the routine switching of the laser helicity and tag the data accordingly.
Next, we will develop offline data analysis of the
asymmetry of back-scattered $\gamma$ photons and fitting method to extract the electron beam polarization.
Real-time measurements of the electron beam polarization will then require automation of the data taking and online analysis procedures.
Reliable beam polarization measurements, once achieved, will enable resonant depolarization experiments 
at BEPC\uppercase\expandafter{\romannumeral 2}, and help enhance the spin physics studies
at the BES\uppercase\expandafter{\romannumeral 3} experiment.

\section{Conclusion}\label{sec6}

Since early 2024, a laser Compton polarimeter has been designed for the electron storage ring of BEPC\uppercase\expandafter{\romannumeral 2},
for vertical polarization measurement via detection of the back-scattered $\gamma$ photons. 
During the long shutdown in 2024,
modifications have been implemented to the beamline and the experimental hutch of a dismantled wiggler source,
that facilitates the delivery of both laser and backscattered $\gamma$ beams through the beamline.
A laser optics system has been set up to guide a focused pulsed laser beam with switchable circular polarization to collide with the electron beam in the storage ring. In July 2025,
the back-scattered
$\gamma$ signals have been successfully detected
with a plastic scintillator, which marks a key milestone in the development of the laser Compton
polarimeter.

\section*{Acknowledgements}
The authors are grateful to the BEPCII and BSRF team for the strong support. Z. Duan acknowledges the illuminating suggestions of V.~V.~Kaminskiy, I.~B.~Nikolaev and S.~A.~Nikitin
on Compton polarimetry, and the
instructive discussions of
Jinguang Wang and Hang Xu on laser
systems.
The permission of M.~N.~Achasov and N.~Yu.~Muchnoi for borrowing the laser-to-vacuum insertion is appreciated.
This study is supported by Center for High Energy Physics, Henan Academy of Sciences, 
National Natural Science Foundation of China (Grant No. 12275283) 
and Youth Innovation Promotion Association CAS (No. 2021012).

\section*{Declarations}

On behalf of all authors, the corresponding author states that there is no conflict of interest.

\bibliographystyle{unsrt}
\bibliography{RDTM}  

\end{document}